# The Bio-habitable Zone and atmospheric properties for Planets of Red Dwarfs

A. Wandel[1] and J. Gale[2]

[1]The Racah Institute of Physics and [2]The Institute of Life Sciences

The Hebrew University of Jerusalem, 91904, Jerusalem, Israel

[1]amri@mail.huji.ac.il

**Abstract**

The Kepler data show that habitable small planets orbiting Red Dwarf stars (RDs) are abundant, and hence might be promising targets to look at for biomarkers and life. Planets orbiting within the Habitable Zone of RDs are close enough to be tidally locked. Some recent works have cast doubt on the ability of planets orbiting RDs to support life.

In contrast, it is shown that temperatures suitable for liquid water and even for organic molecules may exist on tidally locked planets of RDs for a wide range of atmospheres. We chart the surface temperature distribution as a function of the irradiation, greenhouse factor and heat circulation. The habitability boundaries and their dependence on the atmospheric properties are derived. Extending our previous analyses of tidally locked planets, we find that tidally locked as well as synchronous (not completely locked) planets of RDs and K-type stars may support life, for a wider range of orbital distance and atmospheric conditions than previously thought.

In particular, it is argued that life clement environments may be possible on tidally locked and synchronously orbiting planets of RDs and K-type stars, with conditions supporting Oxygenic Photosynthesis, which on Earth was a key to Complex life. Different climate projections and the biological significance of tidal locking on putative complex life are reviewed. We show that when the effect of *continuous* radiation is taken into account, the Photo-synthetically Active Radiation (PAR) available on tidally locked planets, even of RDs, could produce a high Potential Plant Productivity, in analogy to mid-summer growth at high latitudes on Earth.

Awaiting the findings of TESS and JWST, we discuss the implications of the above arguments to the detection of biomarkers such as liquid water and oxygen, as well as to the abundance of biotic planets and life.

1. **Introduction**

Many of the recently discovered habitable exoplanets orbit Red Dwarf stars (RDs), which constitute about 75% of the stars in the Milky Way galaxy. They are characterized by a luminosity much lower than that of our sun. Consequently, habitable planets (in the Habitable Zone, HZ, defined as enabling surface liquid water in the presence of an adequate atmosphere) orbit closely and hence are likely to be gravitationally locked. The findings of the Kepler mission have shown that 10-75% of the RDs have habitable Earth to Super-Earth sized planets, depending on the precise definition of the HZ (e.g. Batalha *et al.* 2013; Dressing and Charbonneau, 2015). This implies that such planets could be found within less than 10 light years from Earth (Wandel, 2015; 2017).

While some authors point out some disadvantages for the evolution of life on planets orbiting M-type stars, such as high levels of XUV radiation and stellar winds, which may cause atmosphere erosion (Heller, Leconte and Barnes 2011; Leconte, Wu, Menou and Murray 2015, Lingam and Loeb 2017, Tilley et al., 2019), other works argue that such planets may nevertheless host life as we know it (e.g. Tarter *et al.*, 2007, Gale and Wandel 2017). They point out that the impact of the high XUV flux received by planets orbiting in the habitable zone of young M-type hosts may be less important for aquatic life. Moreover, atmospheric erosion may be inhibited by a planetary magnetic field (like in the case of Earth). Even for planets with a weak magnetic field, the eroded atmosphere may be compensated by accretion of a secondary atmosphere during the later calmer phases in the evolution of the M-type host, which also last longer. A massive prime atmosphere could survive the extended erosion during the energetic early evolution of M-dwarfs (e.g. Tian, 2009). Furthermore, HZ-planets of K-type stars (which do not have high levels of energetic radiation, characterizing the early evolutionary stages of M-type stars) may be locked or in nearly synchronous rotation, like Venus. Locked or nearly locked planets have been shown to have peculiar climates which may affect their habitability (e.g. Kopparapu et al. 2016; Checlair, Menou, and Abbot 2017). Simulations show that locked or synchronous habitable-zone planets of M-type stars may support liquid water oceans (e.g. Del Genio et al 2019). We look at the habitability of locked, synchronous and slowly rotating planets of M- and K-stars, from the viewpoint of bio-habitability. Using a simple analytic climate-model we derive expressions for the surface temperature dependence on heat circulation and greenhouse heating. A self-consistent analytic treatment of habitability in combination with atmospheric properties is presented. We consider the implications to the evolution and sustainability of life, and in particular oxygenic photosynthesis, an essential precursor of complex life on Earth.

## 2. Model Predictions of the Climates of RD- planets

A major requirement for the appearance and evolution of life is a temperature range which supports liquid water and complex organic molecules. This depends not only on the irradiation from the host star, but to a large extent also on the planet's atmosphere. Global Circulation Models (GCMs) using radiative transfer, turbulence, convection and volatile phase changes can be used to calculate the conditions on planets, given the properties of their atmospheres. Such 3D climate models of M-dwarf planets suggest the presence of liquid water for a variety of atmospheric conditions (e.g. Pierrehumbert 2011, Wordsworth 2015). Climate modeling studies have shown that an atmosphere only 10% of the mass of Earth's atmosphere can transport heat from the day side to the night side of tidally locked planets, enough to prevent atmospheric collapse by condensation (Joshi et al. 1997; Tarter et al. 2007; Scalo et al. 2007; Heng and Kopparla 2012). On locked planets the water may be trapped on the night side (e.g. Leconte 2013), but on planets with enough water or geothermal heat, part of the water remains liquid (Yang et al. 2014). 3D GCM simulations of planets in the habitable zone of M-dwarfs support scenarios with surface water and moderate temperatures (Yang et al. 2014; Leconte et al. 2015; Owen and Mohanty 2016; Turbet *et al.* 2016; Kopparapu et al. 2016; Wolf 2017 to name a few).

While rocky planets with no or little atmosphere, like Mercury, have an extremely high day-night contrast and planets with a thick, Venus-like atmosphere tend to be nearly isothermal, intermediate cases, with up to 10 bar atmospheres, conserve significant surface temperature gradients (e.g. Selsis et al. 2011).

We argue that life clement environments may be supported on tidally locked planets (TLPs). Recent calculations (Wandel 2018; Wandel and Tal-Or 2019) suggest that planets orbiting M-type stars may have life supporting temperatures, at least on part of their surface, for a wide range of atmospheric properties. We extend this treatment to habitable planets of early M-type and K-type stars that are not totally locked but only synchronously rotating, and explore the conditions allowing photosynthesis. Given the very large number of potential planets, and the possibility of liquid water on at least some of the TLPs, we discuss here characteristics of complex life which may evolve on the TLPs and how they may differ from life on Earth.

Wandel (2018) supplements the usual treatment of the habitable zone (the region around a star where planets can support liquid water, e.g. Kasting, Whitmire and Reynolds, 1993), with the definition of a bio-habitable zone, where surface temperatures can support

4complex organic molecules, in addition to liquid water. The atmospheric properties that affect the surface temperature distribution are reduced to three factors: irradiation by the host star, circulative heat redistribution and atmospheric heating due to the greenhouse effect. In the following the presence of surface liquid water and life supporting temperatures are analyzed within this parameter space.

### 3. Surface Temperature distribution on locked planets

The planet is defined as being within the bio-Habitable Zone, if its highest surface temperature is above freezing, that is, $T_{max}$>273K, and its lowest surface temperature is below the highest temperature which allows complex organics, that is $T_{min}$ ~< 400K.

The lower temperature limit is naturally chosen as the freezing point of water, which is only weakly dependent on pressure.

The surface temperature of a planet with no atmosphere is given by

$$\sigma T^4(\theta) = (1 - A)SF(\theta) \qquad (1)$$

where $\sigma$ is the Stefan Boltzmann constant, $A$ is the Bond albedo and $F$ is a function of the angular distance $\theta$ from the sub-stellar point ($\theta$=0). The irradiation or insolation is given by $S=L_*/4\pi a^2$, where $a$ is the distance of the planet from its host star and $L_*$ is host's luminosity.

If the planet is locked and has no horizontal heat circulation, then $F= \cos(\theta)$ for $0<\theta<90$ and $F=0$ for $90<\theta<180$.

On the other hand, planets with an efficient surface heat distribution or rapidly spinning planets are nearly isothermal. *In that case*[1] *F=1/4* and the equilibrium surface temperature is

$$T_{eq} = [(1 - A)S/4\sigma]^{1/4}. \qquad (2)$$

We present a parametric form of the effect of circulative heat redistribution by the atmosphere or an ocean (e. g. Del Genio et al 2019). This form can be enhanced to include the movement of the sub-stellar point due to a spin-orbit resonance, between a completely locked planet with no atmosphere on the one hand and an isothermal planet on the other. We assume that a fixed fraction, *f*, of the stellar irradiation at each point is redistributed homogeneously over the whole planetary surface, due to circulation in the planet's atmosphere or oceans. While in the day hemisphere the heating is modulated by the angular distance ("latitude") from the sub-stellar point, on the night hemisphere it is homogenous:

$$F(\theta) = \begin{cases} \frac{f}{4} + (1-f)\cos(\theta) & 0 \leq \theta \leq 90° \\ \frac{f}{4} & 90° < \theta \leq 180° \end{cases} \quad (3)$$

---

[1] This (1:4) is the ratio between the area of the planet's cross section (receiving the radiation from the host star) and the surface area of the planet, from which the heat is radiated.

The highest surface temperature on TLPs occurs at the sub-stellar point, $\theta=0$. From eq. 1 we have

$$T_0 \equiv T(\theta = 0) = \left[\frac{(1-A)SF(\theta)}{\sigma}\right]^{1/4} \quad (4)$$

The lowest surface temperature occurs at the far end of the night hemisphere, the opposite side of the sub-stellar point, $\theta=180°$.

In the presence of an atmosphere, the temperature is modified by atmospheric screening, namely the fraction of radiation from the host star that reaches the surface, $\alpha$, and atmospheric blanketing, described by the greenhouse factor $H_g$, or the amount by which the atmosphere reduces the heat radiation from the surface via scattering and absorption/reemission). Including these effects eq. 4 becomes

$$T_0 = \left[\sigma^{-1}(1-A)H_g\alpha S(1 - \tfrac{3}{4}f)\right]^{\frac{1}{4}}, \quad (5)$$

while at the opposite point

$$T(\theta=180°) = \left[(1-A)H_g\alpha Sf/4\sigma\right]^{\frac{1}{4}}. \quad (6)$$

### 4. The bio-habitable zone

The conventional habitable zone is defined as the annular region around a star where liquid water can exist on the planetary surface[2]. The inner edge of the habitable zone is located at the distance from the host star, where surface water is completely vaporized or at which water reaches the upper atmosphere, where it can be dissociated by ultraviolet radiation. Runaway greenhouse processes may push this inner edge outwards, while cloud coverage may push it inwards to radii as small as about 0.5 AU for a Sun-like host (e.g., Yang et al. 2014). The extent of the liquid water zone around the host star may be described by the Kombayashi–Ingersoll radiation limit (e.g. Kasting et al. 1993), also called the "runaway greenhouse limit". The outer edge of the habitable zone is determined by water being completely frozen on the planet surface. For planets with a present Earth atmosphere and Sun-like host, this limit may be not much larger than Earth's orbit, because of the positive feedback of the runaway snowball effect. The outer limit may, however, be increased considerably, beyond 2 AU for a Sun-like host, by increasing the $CO_2$ atmospheric abundance (Forget 2013). The freezing point of water may also be lowered by high salinity. Wolf et al. (2017) find four stable climate states defined by their global mean surface temperatures ($T_s$); snowball ($T_s \leq 235$ K), water belt (235 K $\leq T_s \leq$ 250 K), temperate (275 K $\leq T_s \leq$ 315 K), and moist greenhouse ($T_s \geq 330$ K), with the latter three being able to maintain habitable ocean worlds.

---

[2] subject to the presence of a suitable atmosphere

We follow Wandel (2018), defining the habitable zone in the parameter space spanned by the heating factor (*H*) and heat redistribution (*f*). We define the "Bio-habitable zone", which differs from the usual definition of the Habitable Zone in two ways. First, rather than quantifying the spatial boundaries where planets can globally support liquid water, we consider the domain in the parameter space, defined by atmospheric heating and heat redistribution, where liquid water and life-supporting temperatures occur at least on part of the surface of a tidally locked planet. The definition of the bio-habitable zone is different from the traditional habitable zone also in that it excludes high temperatures (T>400K) where water may be liquid, yet is unfavorable for organic life. Depending on the pressure, water may be liquid at temperatures far beyond the temperatures allowing organic complex molecules. We adopt 373K as an upper temperature, although this value may be too low, as organic molecules and even life may survive at somewhat higher temperatures, as demonstrated, e.g. by life forms found near thermal vents in the bottom of Earth's oceans. On the other hand, moist greenhouse runaway may start at average global surface temperatures as low as 355 K (Wolf et al. 2017). The lower temperature limit is chosen as the freezing point of water, which is only weakly dependent on pressure, and is widely accepted as a lower limit for organic processes. Also this limit is somewhat conservative, as high salinity can lower the freezing point to ~250K. In any case, our results are not very sensitive to minor variations of the temperature range. Yet lower surface temperatures, far beyond the Habitable Zone, do not exclude liquid water and life under an ice cover, e.g. in the case of geothermal or tidal heat as in Europa and Enceladus. Tidal heating may be important also in habitable planets of M-type stars (Dobos, Heller and Turner 2017).

## 5. The bio-habitable parameter range

In this section we identify the range of atmospheric parameters suitable for liquid water and complex organic molecules. It is convenient to combine the albedo, the atmospheric screening $\alpha$ and greenhouse factor $H_g$ with the insolation (irradiation) into a single dimensionless parameter, which we call the heating factor

$$H = (1 - A)H_g \alpha \, S/S_\oplus. \tag{7}$$

where $S/S_\oplus$ is the insolation relative to Earth. With this notation eqs. 5 and 6, which give the highest and lowest surface temperature, respectively, can be written as

$$T_{max} = T_0 = 394 \, H^{1/4} (1-\tfrac{3}{4}f)^{1/4} \text{ K} \quad \text{and} \tag{8}$$

$$T_{min} = 278 \, (Hf)^{1/4} \text{ K}. \tag{9}$$

Surface temperature profiles for various values of the combined heating factor $H$ are shown in Fig 1, where a moderate local heat transport (Wandel 2018) has been assumed.

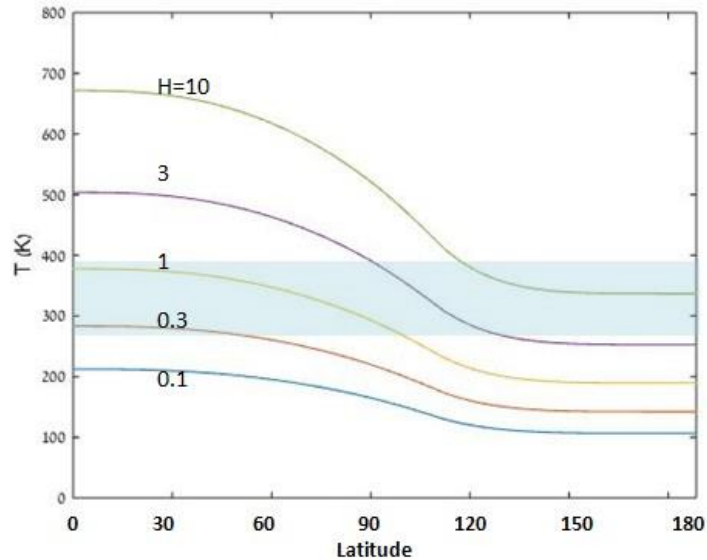

Fig 1. Temperature profiles for several values of the heating factor $H$. A global heat redistribution of $f=0.2$ is assumed. The bio-habitable temperature range is indicated by the shaded light blue area.

The dependence of the minimum and maximum temperatures (eqs. 8-9) on the heating factor $H$, is shown in Fig 2. From the intersection of the curves with the boundaries of the bio-habitable temperature zone it is possible to derive the range of the heating factor. As shown in the figure by the arrows at the bottom, for $273<T<373$K and $f=0.1$, the bio-habitable range of the heating factor turns out to be *0.2<H<30*. Letters denote the terrestrial planets of the Solar System. Note that Venus, Earth and Mars have average temperatures consistent with the equilibrium solid green curve (isothermal surface, *f=1*), while Mercury's temperature gradient is shown by the dashed vertical arrow, indicating that it's effective redistribution parameter *is f<0.1*.

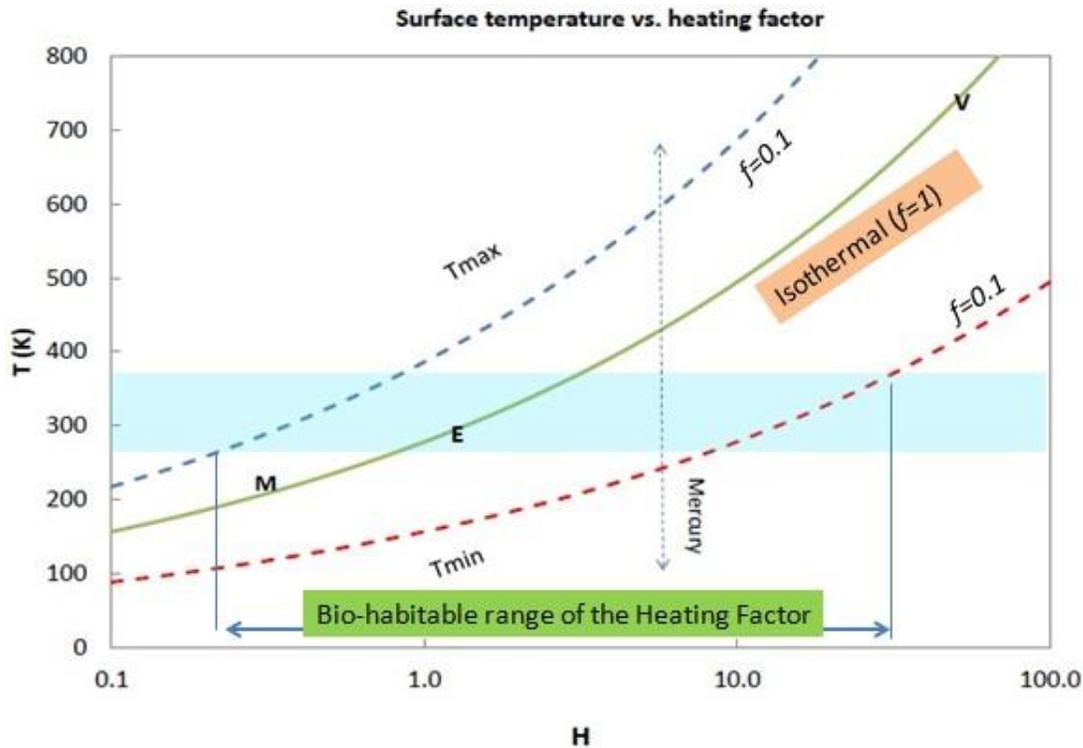

Fig. 2. Minimum and maximum surface temperatures of TLPs as a function of the heating factor *H*. The green solid curve shows the equilibrium temperature for an isothermal planet, while the dashed curves show the highest and lowest surface temperatures of a tidally locked planet with 10% redistribution (*f=0.1*). Also shown are the locations of the four terrestrial planets of our Solar System.

Fig 3 shows the bio-habitable range in terms of the atmospheric heating factor as a function of the distance from the host star. The curves are calculated for a low mass RD with a luminosity of $10^{-4}$ that of our Sun (similar to that of Trappist-1), for several values of the redistribution parameter *f*. In particular, *f=1* corresponds to the isothermal case, which includes rapidly spinning planets. The locations of Proxima Cen b and of three of the Trappist-1 planets are marked on the x-axis. The y-axis shows the atmospheric part of the heating factor (eq. 7) normalized at zero, that is, $(1 - A)H_g\alpha - 1$. The bio-habitability range of the heating factor, which corresponds to the highest and lowest surface temperature being consistent with liquid water at some point on the planet's surface, is given by the vertical distance between the upper and lower curves, corresponding to the redistribution parameter and orbital distance of the planet.

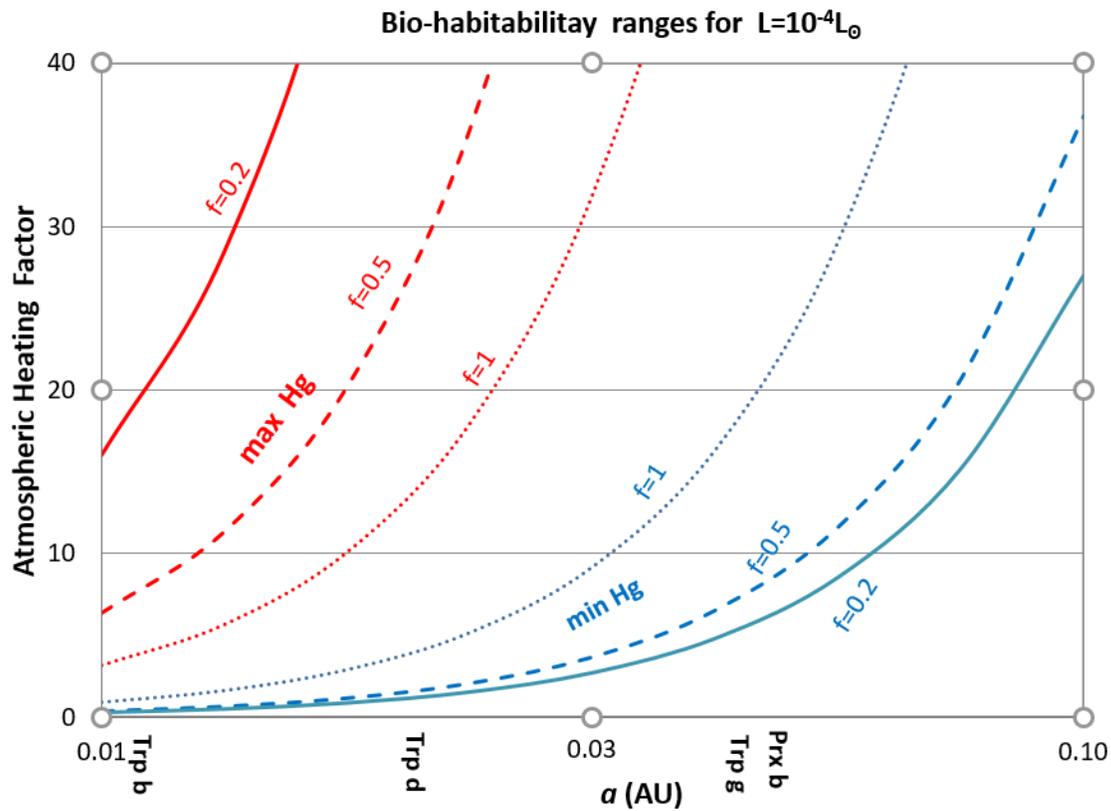

Fig. 3 Maximal (red, upper) and minimal (blue, lower) bio-habitable zone boundaries of the heating factor vs. distance from the host star, for three values of the heat redistribution parameter: $f$=0.2 (solid), 0.5 (dashed) and 1 (dotted). The host luminosity is assumed to be $L=10^{-4}L_\odot$. The locations of Proxima b and 3 of the Trappist-1 planets are marked on the x-axis

## 6. Synchronous orbits and slowly spinning planets

In this section we extend the surface temperature model for tidally locked planets to synchronously orbiting planets (like Mercury, which has a spin-to-orbit period ratio of 3:2) and nearly synchronous ones (e.g. Venus), for which the spin (day/night) period is relatively long, comparable to the orbital period.

If the planet is not completely locked, but rather in a synchronous or nearly synchronous orbit, the sub-stellar point moves slowly. This movement causes various parts of the planet to be illuminated in a slow periodic pattern. This may affect the surface temperature distribution by enhancing the heat redistribution. The importance of this effect depends on the rate of surface temperature change due to heating by radiation from the host star, or cooling by heat emission from the planetary surface into space. Given the average specific heat capacity of the surface and the flux from the host, it is possible to calculate the effective heating and cooling time constants $t_H$, and $t_C$.
The heating timescale is

$$t_H \sim \frac{k \rho \Delta h \Delta T}{(1-A)\alpha S}, \qquad (10)$$

where $\kappa$ is the typical specific heat capacity, $\rho$ – the density, $\Delta h$ – the thickness of the layer that needs to be heated, $\Delta T$ – the temperature difference, $S$ – the radiative flux from the host star and $\alpha$ – the atmospheric screening factor.

As an example, suppose that the surface is rock (typical specific heat capacity of $\kappa = 800$ J/Kg-K and density $\rho \sim 3000$ Kg/m$^3$) and that a layer of $\Delta h \sim 10$ cm needs to be heated, to increase its temperature by $\Delta T \sim 100$K. We normalize the irradiation to that of Earth ($S_\oplus = 1670$ J/m$^2$/sec), which is approximately the insolation within the HZ. Then the heating time required is approximately

$$t_H \sim \kappa \rho \Delta h \Delta T / S \sim 800 \cdot 3000 \cdot 0.2 \cdot 100 \, (S/S_\oplus)^{-1}/1670 \sim 0.17 \, (S/S_\oplus)^{-1} \text{ days.}$$

Taking $A = 0.3$ for the albedo of rock, and assuming $\alpha = 0.7$, the heating time would be approximately 0.33 days. In the case of water, the time is considerably longer. The specific latent heats of melting and evaporation are 334 and 2260 kJ/Kg, respectively, and heating from 273 to 373K would require 420 kJ/Kg, together $\kappa_W \sim 3000$ kJ/Kg. For a 10 cm thick layer of water, melting, heating to the boiling temperature and evaporation takes

$$t_H \sim \kappa_W \rho \Delta h / (S/S_\oplus) \alpha (1-A) \sim 3 \cdot 10^6 \cdot 1000 \cdot 0.1 (S/S_\oplus)^{-1}/1670 \cdot 0.5 \cdot 0.5 \sim 8.8 \, (S/S_\oplus)^{-1} \text{ days,}$$

where we assumed a higher albedo, $A = 0.5$ for ice/water and a lower $\alpha = 0.5$ (enhanced screening) to take into account water vapor (clouds).

Similarly, assuming the surface heat loss rate is approximately black body, moderated by the atmospheric greenhouse factor $H_g$, the cooling from the high daytime temperature to the low nighttime one is

$$t_C \sim \frac{k \rho \Delta h \Delta T H_g}{\alpha \sigma T^4} \qquad (11)$$

Assuming the temperature is equal to the equilibrium surface temperature, substituting eq. 2 and adding the atmospheric effects (screening and greenhouse) gives

$$t_C \sim \frac{4k\,\rho\,\Delta h\,\Delta T H_g}{\alpha(1-A)H_g F} = 4t_H. \tag{12}$$

For $T=T_0$ rather than $T_{eq}$, eq. 6 gives $t_C\sim(4-3f)t_H$. For the parameter values of the above example eqs. 11-12 give $t_H\sim1.4(S/S_\oplus)^{-1}$ days for rock and $t_C\sim33(S/S_\oplus)^{-1}$ days for water. Coincidently, this cooling time range is comparable to the orbital period in the habitable zone of M-type stars.

In order to estimate the effect of rotation on the surface temperature, the heating and cooling times have to be compared with the rotation period. For synchronous or nearly synchronous orbits we define an effective rotation time $t_r$ as the time it takes for the host star, as seen from a fixed point on the planet, to complete one cycle ("day")[3]. Obviously, if $t_r$ is much longer than the heating/cooling times, the momentary surface temperature distribution of the planet can be approximated by that of a TLP. In the opposite case, when $t_r\sim<t_H$ or $t_r\sim<t_C$, the surface does not have enough time to adjust to the TLP temperature, hence the surface temperature varies between the isothermal equilibrium temperature $T_{eq} = 278\,H^{1/4}$ K and the TLP temperature given by eqs. 8-9.

### 7. Life on Tidally Locked Planets

Although the climates on TLPs would not have the same seasonal and circadian cycles of radiation as on Earth, there may be a slow variation, e.g. in near synchronously, slowly orbiting planets, like Venus, if the axis of the planet was tilted to the ecliptic (Kite et al, 2011). However, we assume here that although such variations are possible, and likely, the environment on many TLPs would be generally stable.

Recently Stevenson (2017, 2018a, 2018b) considered that the surface temperature structure of locked planets would deny the development of biological complexity, as the habitable belt around the sub-stellar point may be narrower than the latitudinal habitable bands on asynchronously rotating planets. However, as can be seen in Fig. 1, the habitable region on a locked planet often extends over a significant part of the surface. Actually, our results support the suggestion (Gale and Wandel 2017), that locked HZ planets provide conditions not very different from those during the summer season at high latitudes on Earth. Furthermore, the time available for the evolution of biological complexity is even longer than on Earth, as the luminosity of the host changes slower for the lower Main Sequence M and K type stars, whose lifetime is much longer than that of the Sun. M type stars have an early flaring phase, with a high luminosity of X and UV radiation, which could endanger life unless it is protected under water. However, K type stars, as well as M type ones in their later evolutionary stages, are more quiescent and hence more hospitable also to land life.

---

[3] For synchronous rotation planets $t_r$ is of the order of the orbital period. E.g. the case of Mercury 's orbital resonance. It takes Mercury about 59 Earth days to spin once on its axis (the rotation period), and about 88 Earth days to complete one orbit about the Sun. However, the length of the day on Mercury (sunrise to sunrise) is 176 Earth days.

In the case of M-type stars, tidally locked planets are mostly within the HZ. Recently, Lingam and Loeb (2017) suggested that planets orbiting red dwarfs are less likely to be habitable due to atmospheric erosion as they may experience greater exposure to stellar winds and EUV Radiation. However, atmospheric erosion may be avoided or limited by several factors, such as a strong enough planetary magnetic field or a thick primary atmosphere and the acquisition of a secondary atmosphere after the early energetic high activity phase of the M-host (Tarter et al 2007, Gale and Wandel 2017).

8. **Importance of a High Oxygen Atmosphere for the evolution of Complex Life**

As depicted in Fig 4, following Lyon et al (2014) and others, life first appeared on Earth some 4 Gy before present (BP). For ~3.5Gy only single celled life prevailed under a very low-oxygen atmosphere[4]. The first anoxygenic photosynthesis appeared about 3.8Gy BP, followed one Gy later by water splitting, oxygenic photosynthesis. The release of oxygen from water resulted in an increase in the level of $O_2$ in the atmosphere. The first significant $O_2$ rise began in ~2.0Gy BP when it reached some 10% (the "Great Oxygen Event" or GEO, which was partly responsible for some of the first great Ice Ages). The relatively low initial levels of oxygen delayed the evolution of animal life in the Mid-Proterozoic (2.5 to 0.6 Gy BP) (Planevsky et al 2014). Around 0.6Gy BP the atmospheric level started rising more steeply, reaching the present level in ~500 My BP, During the same period, first Protists (unicellular Eukaryotes) then complex, multi-cellular Eukaryotes evolved. This has been termed the Cambrian "Explosion", although it is now recognized that the process took many tens of millions of years.

---

[4] As compared to the present abundance of 21%

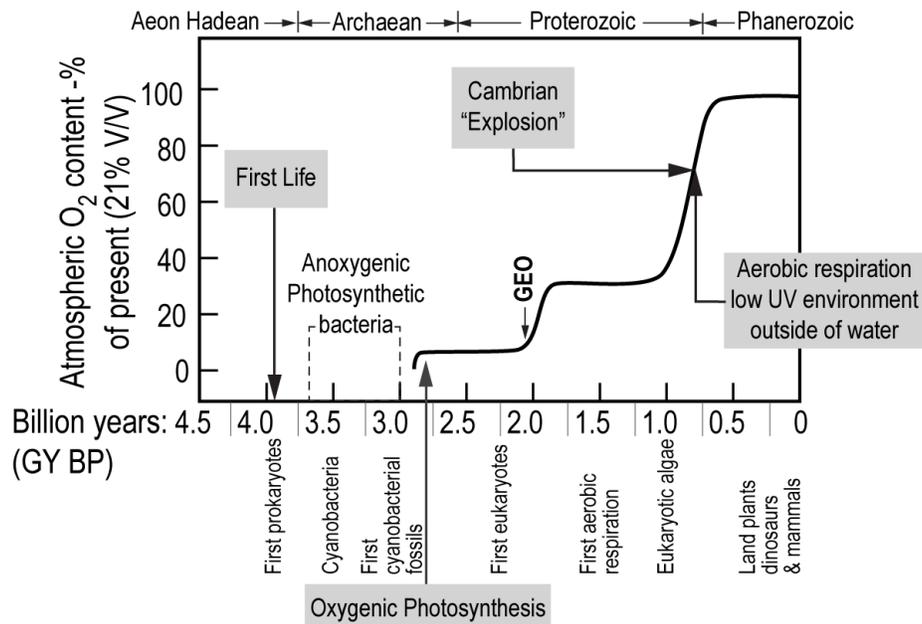

Fig 4. The evolution and consequences for Life of the high Oxygen atmosphere on Earth.

The reason for the coincidence of the high oxygen atmosphere and the appearance of complex life has been ascribed to two effects of high oxygen: the enablement of aerobic respiration, which produces nine times the energy from each unit of substrate than anaerobic respiration and the screening of UV radiation, which allowed the emergence of life from the protection of water, to dry land (Lambers et al 2008). A rapid supply of energy is required for complex life forms, especially for motility and brain metabolism (Belanger et al 2011). On Earth, dry-land life is protected by the UV absorption of oxygen and the ozone produced. However, this consideration may be less important on planets of M-type stars after their early flary stage, as Red Dwarfs radiate much less than our Sun in the UV waveband.

9. Could TLPs support oxygenic photosynthesis and oxygen-rich atmospheres?

The surface temperatures of Red Dwarf stars are cooler than our sun (2,500 – 4,000 vs 5,800°K). Although they radiate less photons in the Photo-synthetically Active Radiation (PAR) waveband (400-700nm) than is incident on Earth, there is still enough PAR flux to support Earth-like plant life (Gale and Wandel, 2017).

Ritchie et al. (2018) made theoretical estimates and laboratory measurements of the photosynthesis of different species of algae exposed to radiation levels which could be expected at the Sub-stellar points of TLPs. They concluded that the PAR incident on TLPs may produce high levels of basic plant productivity: from 13-22% of what would be expected on Earth. Their calculations were made for productivity per second or per hour. However, compared to the annual incident radiation on Earth, this may be an underestimate. As pointed out by Gale and Wandel (2017), vegetation on TLPs would be exposed to continuous light. This is analogous to the summer season at high Earth latitudes, where plants receive almost continuous radiation, resulting in lush growth. On most TLPs there will be no seasons and radiation would be continuous.

By way of example, the vegetation in midsummer on the Scottish Loch Creran, at latitude 57.5°N enjoys 17 hours of moderate PAR (Johnston et al (1977). "Enjoys" as radiation at this latitude does not reach damaging levels (Gale and Wandel, 2017, Fig.5).

Zhang et al. (2017) calculated the potential Gross Primary Production of plants (GPP) as a function of PAR at different latitudes on Earth. Maximum GPP at 55°N was found to be very high (in midsummer) resulting in an annual GPP (almost entirely produced in the summer months) reaching to about one third of that in the tropics (0±10° latitude). By extrapolation, this should at least double the productivity on TLPs as calculated by Ritchie et al. Furthermore, damaging levels of radiation would be avoided on TLPs, by the plants "selecting" locations with optimum radiation regimes, between the Sub-stellar point and light termination (Gale and Wandel 2017).

Some Earth algae already have pigments which utilize small parts of the NIR (Larkum 2018, Nurnberg et al, 2018). King et al, (2007a,b) have discussed the theoretical possibility that plants exposed to the copious Near Infra-Red Radiation (NIR) of M- stars (in the 700-1000 nm waveband) could evolve pigments which would utilize this waveband for Oxygenic Photosynthesis. However, although theoretically possible, there would be little evolutionary pressure for this step to appear, as there may be a surfeit of PAR in the 400-700nm waveband on TLPs. Moreover, photosynthesis in the NIR would have lower photonic efficiency (Wolstencroft and Raven, 2002). On Earth, land plants actually evolved leaves which reject NIR (by reflection), presumably to reduce water loss in situations where there is sufficient PAR (Gale and Wandel, 2017).

In view of the above, it is concluded that there should be no shortage of PAR on TLPs (even before extending the PAR-range to 400-1000nm). If high rates of Oxygenic Photosynthesis produce a high oxygen atmosphere then, in analogy to Earth, conditions would exist for the appearance of complex, multi-cellular, life. However, such life would probably differ from that on Earth which is subject to seasonality and a 24-hour light-dark radiation cycle, which are absent on TLPs. We emphasize again that the mere appearance of clement conditions does not guarantee the evolution of complex or even simple life. By way of analogy, on "Goldilocks" Earth, most geological niches which contain liquid water do not support life (Jones and Lineweaver, 2010).

## 10. The effect of Tidal Locking and the Absence of Seasonal and Circadian cycles on Complex life

Habitable planets of M-stars would be close enough to be tidally locked. We discuss the special conditions on such planets, which are different to those on Earth, and how, after the mono-cellular evolutionary phase, they will affect complex, multi-cellular life. We first address the question of the importance of the oxygen-rich environment to the evolution of complex life on Earth.

Vegetation on Earth operates anabolic and catabolic metabolisms. Photosynthesis, the beginning of an anabolic chain, is active in the light, while respiration, with its basically catabolic chain, followed by the construction of cellular components, continues throughout the day. Enzymes involved in the two processes are balanced accordingly (Amthor, 1989). This would necessarily be different on TLPs, where both operate continuously.

On Earth, vegetation adapts to seasons, from little radiation seasonality in the tropics (with climate affected by seasonal rainfalls) to extreme variation at far northern, or southern latitudes. Plants respond with adaptations of germination, flowering, root versus stem production, leaf fall and many other processes. These temporal responses are governed by mechanisms which measure day/night length and temperature cycles (Lambers et al, 2008). Such adaptations would be absent on TLPs, where plants would be exposed to continuous steady radiation, although geographic features (Stevenson, 2018) and planet obliquity (Kite et al 2011) may produce some climate variations.

Animal life on Earth shows response to seasonality by special adaptations, such as migration, hibernation of large animals during the cold of winter in latitudes above ~ 50°, north or south, and estivation of some drought avoiding animals in desert summers (Storey and Storey, 1990). Such adaptations would not be relevant to life on TLPs.

Earth's movement around the Sun at an oblique angle to the ecliptic, produces seasons at latitudes not close to the equator. However, the most significant movement of Earth, is its rotation. Earth's rotation was not always of 24 hours, but started at 10 hours (Hadhazy 2010). The day slowly increased in length as the moon receded. Hadhazy, who reviewed the literature on Moon – Earth attraction and the many other factors which effect day length, estimated a 22h rotation at 650My BP. Sonett et al (1996) estimated that at 900 My BP, day length was as low as 18 hours.

This means that mechanisms which evolved to govern circadian rhythms must have included resettable clocks. This is something which all who have travelled long distances by plane, East or West, are familiar with. We are at first confused by "Jet lag", but our internal clocks readjust, at the rate of about one day for each hour of change. Such adaptations would be unnecessary and therefor absent on TLPs.

The most outstanding adaptation of animal life to Earth's rotation (above the arthropod and cordate phyla), is sleep during the dark periods of the day (Stickgold and Walker, 2009). Sleep appears throughout the animal world, and is recognized even in some insects such as Drosophila (Shaw, 2000). There are only conjectures for the reason for this phenomenon. Sleep probably evolved to conserve resources during periods when animals could not hunt or gather food and would risk injury and exposure to dark adapted enemies. What we do know today, is that advanced nervous systems have adapted to utilize sleep for maintenance (Krueger et al 2008). Sleep deprivation in humans, and probably in other sleeping animals, causes disorientation and even immune system damage (Immeri and Opp, 2009). This maintenance is probably a secondary adaptation. If periodic sleep does not appear in higher, complex lifeforms on TLPs, some other form of maintenance would probably evolve. An example of such an adaptive mechanism evolved on Earth in Dolphins and Porpoises. These animals keep half their brains awake while the other half sleeps. This allows them to continue eating throughout the day, while being on guard against night- active predators (Makhameter, 1987).

**11. The cosmic abundance of Life**

The above arguments show that life bearing planets of RDs may be copious. Following Wandel (2015, 2018) we estimate the abundance and probability of biotic life. Assuming that biotic life is long lived, as on Earth (~4Gyr for mono-cellular life), the number of biotic planets can be expressed by a Drake-like equation,

$$N_b = N_* F_s F_{EHZ} F_b, \qquad (13)$$

where $N_*$ is the number of stars in the Galaxy, $F_s$ is the fraction of stars suitable for evolution of life and $F_{EHZ}$ is the fraction of such stars that have Earth-size planets within their habitable zone. The last parameter, $F_b$, is the (yet unknown) biotic probability, that a habitable planet actually becomes biotic within a few billion years. In the following we assume that the duration of biotic life, once it has emerged on an Earthlike planet, is like on Earth, at least a few billion years.

Note that this assumption does not refer to any single species, but rather to mono-cellular life, which has existed on Earth for about 4 Gy. In particular, oxygenic photosynthetic organisms, which are responsible for Earth's high Oxygen atmosphere, appeared about 2.8 Gy BP (Fig. 4) and survived ever since.

With this assumption the average distance $d_b$ between biotic neighbor planets can be shown to be (Wandel 2015; 2017)

$$d_b \sim 3 \, (n_* F_s F_{EHZ} F_b)^{-1/3} \text{ pc} \qquad (14)$$

where $n_*$ is the stellar number-density in the solar neighborhood. For Red Dwarf stars, $n_* \sim 0.2$ pc$^{-3}$ and $F_s \sim 0.7$ (Henry et al. 2006; Winters et al. 2015).

The fractional abundance of Earth-size habitable planets, $F_{EHZ}$, can be estimated from the Kepler data (e. g. Dressing and Charbonneau 2015), $F_{EHZ} \sim 0.1$-$0.75$, depending on the definition of the habitable zone. Substituting in eq. 14 gives

$$d_b \sim (0.7\text{-}1.4) \, F_b^{-1/3} \text{ pc.} \qquad (15)$$

For example, assuming one in ten Earth-size habitable planets of M-stars becomes biotic ($F_b = 0.1$), eq. 15 gives $d_b \sim 1.5$-$3$ pc. The results of the previous sections may indicate that the actual value of $F_{EHZ}$ is nearer to unity than to 0.1, and furthermore, increase the expected value of $F_b$.

The expected number of biotic planets within a distance $d$ may be derived from eq. (15) by substituting the values of $n_*$ and $F_s$,

$$N_b(d) \sim 160 \, F_{EHZ} F_b \left(\frac{d}{10 \text{ pc}}\right)^3 \qquad (16)$$

Figure 5 shows the expected number $N_b$ as a function of the distance and the product $F_{EHZ}F_b$. The previous sections imply that the *effective* value of $F_{EHZ}$ (the fraction of Earth-sized planets in the bio-habitable zone) could be close to unity. Bio-habitability, as defined in sections 6-7 (surface temperatures supporting liquid water and organics), allows evolution of Oxygenic Photosynthesis[5] (Gale and Wandel 2017) and hence an oxygenic bio-signature. This gives a direct relation between the observed number of planets with oxygenic signatures and the biotic probability $F_b$, with the caveat of non-biotic false-positive oxygen signatures.

---

[5] For gravitationally locked planets, photosynthesis may be excluded from the night-hemisphere, unless the atmosphere is sufficiently dense as to scatter light from the day hemisphere.

TESS, launched in April 2018, is expected to find hundreds of transiting Earth-to-Super-Earth size planets, many of them in the habitable zones of nearby M-dwarfs (*Ricker, et al. 2014; Seager 2014)*. Given the bio-signature detection range, we can determine the number of expected candidates by calculating the cumulative number of biotic planets as a function of distance (Wandel 2018).

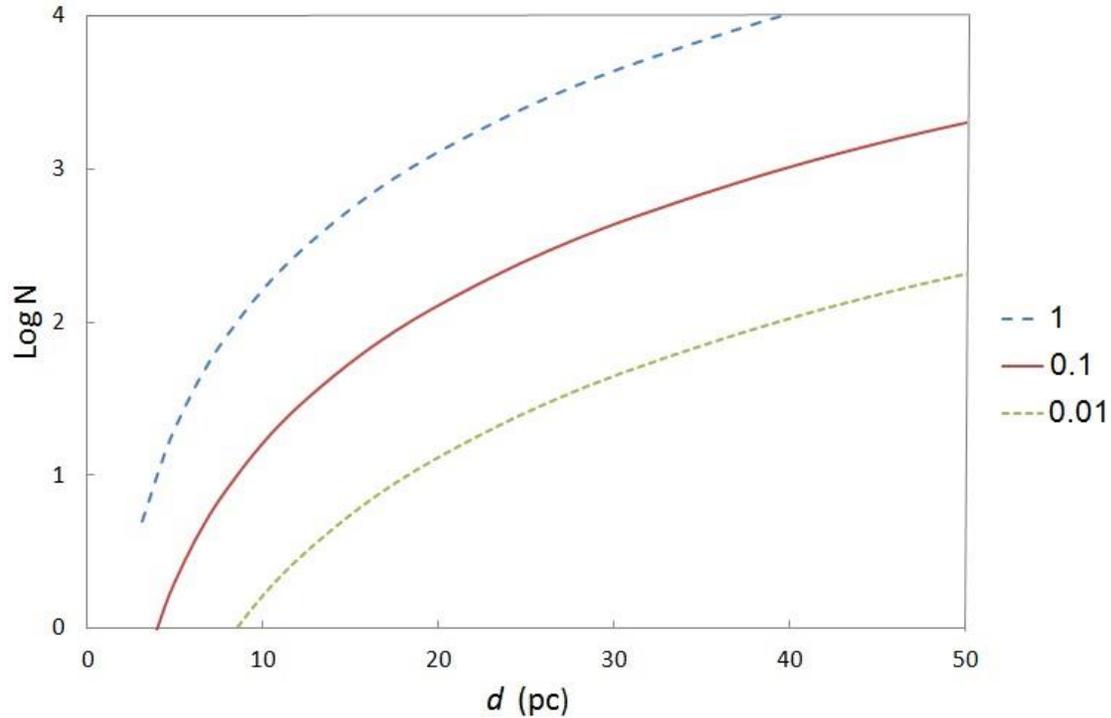

Figure 5. The number of red dwarf planets with oxygenic bio-signatures expected within a distance *d*, for several values of the product $F_{EHZ} F_b$ (marked on the right), where $F_b$ is the biotic probability *(see text)* and $F_{EHZ}$ is the fraction of M-type stars with Earth-sized habitable-zone planets.

Using eq. 16 and Fig 5. we can calculate the expected number of candidate planets within a given distance (the bio-signature detection range). If the M-dwarf habitability fraction is $F_{EHZ}$=0.5 (i.e. 0.5 Earth-sized HZ-planets per Red Dwarf), one could expect ~2000 habitable planets of RDs within 30pc. With a transiting probability of 1-2%, TESS can be expected to detect 20-40 transiting ones.

As an example, let us assume that out of 30 candidates we find oxygenic signatures in 20 planets, and half of them have other supporting bio-signatures. One could then estimate the biotic probability to be $F_b$~0.3±0.05.

Of course, such signatures may be inconclusive, as oxygen may be produced abiotically, but combinations of several components, such as Oxygen, water and Methane could be more unique (e.g. Seger et al 2010).

## 12. Discussion and Conclusions

The habitability of gravitationally locked planets is investigated, taking into account irradiation, albedo, circulative heat transport and atmospheric effects such as screening and greenhouse effect. Our model supplements earlier calculations, combining the circulation and radiative heating calculations with habitability and life-supporting conditions. We find that habitable-zone planets of M-dwarf stars may have temperatures supporting liquid water and complex organic molecules on at least part of their surface, for a wide range of irradiations and atmospheric properties, in particular taking into account the greenhouse and circulation effects.

Some recent works argued that planets of Red Dwarf stars may be less favorable to life because of atmosphere erosion and being tidally locked. Projections of the present model suggest that the Habitable Zones of RD stars may be less hostile to life than previously thought. We argue that Oxygenic Photosynthesis and perhaps complex life on tidally locked and synchronously orbiting planets around Red Dwarf stars may be possible. Combined with the vast number of RD stars and their planets in the Galaxy, these results seem to make the possibility of finding life on such planets plausible.

It is argued that PAR, between 400-700nm would not be lacking on RD planets, even for plants developing under water, where they are protected from the XUV radiation. Moreover, should plant life spread to dry land, after the evolution of an XUV absorbing atmosphere, theoretical considerations indicate that photosynthetic pigments could evolve to utilize the copious Near Infra-Red radiation of the RD stars, which is not being absorbed by a water cover. This may add to the availability of PAR, even though its efficiency would be lower than that in the 400-700nm waveband.

By analogy to Earth, if oxygenic photosynthesis is possible and produces an oxygen-rich atmosphere, then complex life could evolve. However, plants and animals, which evolve on such planets, are likely to differ from those on Earth in many ways, such as the anabolic/catabolic enzyme balance and circadian and seasonal rhythms in plants. Animals would not evince seasonal hibernation and estivation and, not being exposed to diurnal radiation cycles, would have different sleep patterns.

Using Kepler data, we estimate the abundance of habitable candidates to harbor life. E.g. within a biomarker detection range of 30 parsec TESS should find 20-40 transiting habitable super-Earth-sized planets. Observing such a sample of planets and searching for spectral signatures of water and oxygen could yield an estimate of the abundance of biotic planets and of the probability for the evolution of organic life on habitable planets.

**Acknowledgements**

We thank the reviewer for his helpful comments.